\documentclass[twocolumn,showpacs,preprintnumbers,amsmath,amssymb]{revtex4}
\usepackage{graphicx}
\usepackage{dcolumn}
\usepackage{bm}
\usepackage{color}

\begin{document}

\title{Deformed quasiparticle-random-phase approximation
for neutron-rich nuclei using the Skyrme energy density functional
}

\author{Kenichi Yoshida}
\affiliation{Nishina Center for Accelerator-Based Science,
The Institute of Physical and Chemical Research (RIKEN),
Wako, Saitama 351-0198, Japan}
\author{Nguyen Van Giai}
\affiliation{
Institut de Physique Nucl\'eaire, IN$_{2}$P$_{3}$-CNRS,
and Universit\'e Paris-Sud, F-91406 Orsay Cedex, France
}%

\date{\today}

\begin{abstract}
We develop a new framework of
the deformed quasiparticle-random-phase approximation (QRPA)
where the Skyrme density functional and the
density-dependent pairing functional are consistently treated.
Numerical applications are carried out for the isovector dipole and
the isoscalar quadrupole modes in the spherical $^{20}$O and
in the deformed $^{26}$Ne nuclei, and the effect of the
momentum dependent terms of the Skyrme effective interaction for
the energy-weighted sum rule is discussed.
As a further application, we present for the first time the
moments of inertia of $^{34}$Mg and $^{36}$Mg using the
Thouless-Valatin procedure based on the self-consistent deformed QRPA,
and show the applicability of our new calculation scheme not only for the vibrational modes
but also for the rotational modes in neutron-rich nuclei.
\end{abstract}

\maketitle

\section{Introduction}
Exploring nuclei far from the stability line is one of the
most actively studied fields in nuclear physics.
The exotic nuclei have revealed many features of atomic nuclei that are quite different
from stable nuclei.
Examples are the emergence of the neutron halo~\cite{tan85} and the skin~\cite{suz95} structures,
the soft dipole excitations~\cite{sac93},
the modifications of some magic numbers~\cite{iwa00,gui84}
and the appearance of new magic numbers instead~\cite{oza00},
the onset of new regions of deformation~\cite{mot95}.
These new features are strongly connected with the presence of the loosely bound
neutrons and the coupling with the positive energy continuum states.
Under the new environment, the nuclear many-body
correlations such as the pairing and the deformation could have
also unique features in exotic nuclei~\cite{dob84,dob96,ben99,ben00b,mis97,ham04a,yos05a,ham04,ham05}.

For describing multipole responses in exotic nuclei in the region of medium-heavy systems,
there have been many attempts employing the particle-hole random phase approximation
(RPA)~\cite{ham96,ham99,shl03}
or the quasiparticle-RPA (QRPA)~\cite{mat01,mat05,miz07,hag01,
kha02,yam04,ter05,ter06,vre01,paa03,paa04,paa05,cao05,gia03,sar04,per05}
on top of the self-consistent Hartree-Fock (HF) or
Hartree-Fock-Bogoliubov (HFB) mean fields.
(See Refs.~\cite{ben03,vre05,paa07} for
extensive lists of references concerning the self-consistent
(Q)RPA and mean-field calculations.)
These studies are largely restricted to spherical systems.
Recently, low-lying RPA modes in deformed
neutron-rich nuclei have been investigated by several
groups~\cite{yos05,nak05,ina06,sar98,mor06,urk01,hag04a,pen07}.
These calculations, however, do not take into account the pairing
correlations, or rely on the BCS approximation for pairing, which is inappropriate for
describing the pairing correlations in drip line nuclei due to the
unphysical nucleon gas problem~\cite{dob84}.

In order to discuss simultaneously effects of nuclear deformation
and pairing correlations including the unbound quasiparticle states,
we have developed in Ref.~\cite{yos06} a calculation
scheme that carried out the deformed QRPA calculation based on the
coordinate-space HFB formalism.
The residual interaction in the particle-hole (p-h) channel was a simplified Skyrme interaction
neglecting the momentum dependence~\cite{ber91}
while a deformed Woods-Saxon potential was employed for the mean field.
In Ref.~\cite{yos08}, one step further was accomplished by using a self-consistent Skyrme-HFB
deformed mean field while the p-h residual interaction corresponding to the same
Skyrme force was approximated by its Landau-Migdal (LM) form~\cite{bac75,gia81}.
However, the resulting energy-weighted sum rule (EWSR) for the
isovector dipole response was not fulfilled very accurately.

A full consistency between  the QRPA and HFB calculations is
required for a quantitative description of the multipole strengths in exotic nuclei.
This means that the same effective interaction
or the same energy density functional is used for both calculations.
We note that fully consistent HFB+QRPA calculations with the Gogny effective interaction for
deformed nuclei are now available~\cite{per08}, but the use of a harmonic oscillator basis is a drawback
for describing the unique spatial structure of quasiparticle wave
functions near the Fermi level in neutron drip-line nuclei.

We therefore develop in this article a new method
for solving the Skyrme-HFB-QRPA problem in deformed systems while
keeping the full velocity dependence of the p-h residual interaction.
The Skyrme-HFB mean field is calculated in the
coordinate-space representation.
Numerical calculations are
performed in order to investigate the effects of the explicit
treatment of the momentum-dependent terms of the effective interaction on the EWSR
of multipole responses and on the moments of inertia of deformed
neutron-rich nuclei.
The decoupling between the spurious mode of
translation and intrinsic excitations is reasonably obtained in these
consistent calculations.

The article is organized as follows:
In Sec.~\ref{method}, the method is explained.
In Sec.~\ref{results}, we perform the numerical calculations and investigate properties of the
isovector/isoscalar dipole and the isoscalar quadrupole modes in
$^{20}$O, the isovector/isoscalar dipole modes in $^{26}$Ne, and
the moments of inertia of $^{34}$Mg and $^{36}$Mg as well as the
isoscalar quadrupole mode.
Sec.~\ref{summary} contains the conclusions.
We summarize in the Appendix the main formulas relevant to the Skyrme energy
density functional to show clearly what are the terms that are
included in our approach.

\section{\label{method}Method}
In the present section, we explain our method of the deformed QRPA based on
the Skyrme density functional.
We solve the HFB equations~\cite{dob84,bul80}
\begin{align}
\begin{pmatrix}
h^{q}(\boldsymbol{r},\sigma)-\lambda^{q} & \tilde{h}^{q}(\boldsymbol{r},\sigma) \\
\tilde{h}^{q}(\boldsymbol{r},\sigma) & -(h^{q}(\boldsymbol{r},\sigma)-\lambda^{q})
\end{pmatrix}
\begin{pmatrix}
\varphi^{q}_{1,\alpha}(\boldsymbol{r},\sigma) \\
\varphi^{q}_{2,\alpha}(\boldsymbol{r},\sigma)
\end{pmatrix} \notag \\
= E_{\alpha}
\begin{pmatrix}
\varphi^{q}_{1,\alpha}(\boldsymbol{r},\sigma) \\
\varphi^{q}_{2,\alpha}(\boldsymbol{r},\sigma)
\end{pmatrix} \label{HFB_equation}
\end{align}
in coordinate space using cylindrical coordinates $\boldsymbol{r}=(\rho,z,\phi)$.
We assume axial and reflection symmetries.
Here, $q=\nu$ (neutron) or $\pi$ (proton).
For the mean-field Hamiltonian $h$, we employ the SkM* interaction~\cite{bar82}
in the present numerical applications.
Details for expressing the densities and currents in the cylindrical coordinate
representation can be found in Refs.~\cite{ter03,sto05}.
The pairing field is treated by using the density-dependent contact
interaction~\cite{cha76},
\begin{equation}
v_{pair}(\boldsymbol{r},\boldsymbol{r}^{\prime})=\dfrac{1-P_{\sigma}}{2}
\left[ t_{0}^{\prime}+\dfrac{t_{3}^{\prime}}{6}\varrho_{0}^{\gamma}(\boldsymbol{r}) \right]
\delta(\boldsymbol{r}-\boldsymbol{r}^{\prime}). \label{pair_int}
\end{equation}
where $\varrho_{0}(\boldsymbol{r})$ denotes the isoscalar density
and $P_{\sigma}$ the spin exchange operator.
Assuming time-reversal symmetry and reflection symmetry with respect to the $x-y$ plane,
we have only to solve for positive $\Omega$ and positive $z$.
We use the lattice mesh size $\Delta\rho=\Delta z=0.6$ fm and a box
boundary condition at ($\rho_{\mathrm{max}}=9.9$ fm, $z_{\mathrm{max}}=9.6$ fm)
for $^{20}$O and $^{26}$Ne,
and ($\rho_{\mathrm{max}}=9.9$ fm, $z_{\mathrm{max}}=12$ fm) for Mg isotopes.
The quasiparticle energy cutoff is chosen at $E_{\mathrm{qp,cut}}=60$ MeV
and the quasiparticle states up to $\Omega^{\pi}=15/2^{\pm}$ are included
(for $^{26}$Ne, we include the quasiparticle states up to $\Omega^{\pi}=13/2^{\pm}$ in order
to compare with the results of Ref.~\cite{yos08}).
Our calculation scheme for solving the HFB equations is quite similar to Ref.~\cite{oba08},
whereas the reflection symmetry was not imposed in Ref.~\cite{oba08}.

Using the quasiparticle basis obtained
as the self-consistent solution of the HFB equations (\ref{HFB_equation}),
we solve the QRPA equation in the matrix
formulation~\cite{row70}
\begin{equation}
\sum_{\gamma \delta}
\begin{pmatrix}
A_{\alpha \beta \gamma \delta} & B_{\alpha \beta \gamma \delta} \\
B_{\alpha \beta \gamma \delta} & A_{\alpha \beta \gamma \delta}
\end{pmatrix}
\begin{pmatrix}
X_{\gamma \delta}^{\lambda} \\ Y_{\gamma \delta}^{\lambda}
\end{pmatrix}
=\hbar \omega_{\lambda}
\begin{pmatrix}
1 & 0 \\ 0 & -1
\end{pmatrix}
\begin{pmatrix}
X_{\alpha \beta}^{\lambda} \\ Y_{\alpha \beta}^{\lambda}
\end{pmatrix} \label{eq:AB1}.
\end{equation}
The residual interaction in the particle-hole (p-h) channel appearing
in the QRPA matrices $A$ and $B$ is
derived from the Skyrme density functional,
\begin{equation}
v_{ph}(\boldsymbol{r},\boldsymbol{r}^{\prime})=\dfrac{\delta^{2} \mathcal{E}_{\mathrm{Sky}}}
{\delta \varrho(\boldsymbol{r}^{\prime}) \delta \varrho(\boldsymbol{r})},
\end{equation}
where we neglect the spin-orbit interaction term
$C_{t}^{\nabla J}$ in Eq.~(\ref{Sky_energy}) as well as the
Coulomb interaction to reduce the computing time.
We also drop the so-called $``{J}^{2}"$ term $C_{t}^{T}$ in both
the HFB and QRPA calculations.
Thus, the residual interaction reads
\begin{align}
v_{ph}(\boldsymbol{r},\boldsymbol{r}^{\prime})&=
(a_{0}+a_{0}^{\prime}\tau\cdot\tau^{\prime}+
(b_{0}+b_{0}^{\prime}\tau\cdot\tau^{\prime})\sigma\cdot\sigma^{\prime})
\delta(\boldsymbol{r}-\boldsymbol{r}^{\prime}) \notag \\
&+(a_{1}+a_{1}^{\prime}\tau\cdot\tau^{\prime}+
(b_{1}+b_{1}^{\prime}\tau\cdot\tau^{\prime})\sigma\cdot\sigma^{\prime}) \notag \\
& \times (\boldsymbol{k}^{\dagger 2}\delta(\boldsymbol{r}-\boldsymbol{r}^{\prime})
+ \delta(\boldsymbol{r}-\boldsymbol{r}^{\prime}) \boldsymbol{k}^{2}) \notag \\
& + (a_{2}+a_{2}^{\prime}\tau\cdot\tau^{\prime}+
(b_{2}+b_{2}^{\prime}\tau\cdot\tau^{\prime})\sigma\cdot\sigma^{\prime}) \notag \\
& \times (\boldsymbol{k}^{\dagger}\cdot\delta(\boldsymbol{r}-\boldsymbol{r}^{\prime})\boldsymbol{k} ),
\label{v_res_ph}
\end{align}
where $\boldsymbol{k}=(\overrightarrow{\nabla}-\overrightarrow{\nabla}^{\prime})/2i$ and
$\boldsymbol{k}^\dagger=-(\overleftarrow{\nabla}-\overleftarrow{\nabla}^{\prime})/2i$.
The coefficients in Eq.~(\ref{v_res_ph}) are given in Ref.~\cite{ter05}
(For simplicity, the coefficients 
$a_{0},a_{0}^{\prime},b_{0}$ and $b_{0}^{\prime}$ here include the
density dependent terms and rearrangement terms of $a_{3}-f_{3}$ in Ref.~\cite{ter05}).
We do not include the pairing rearrangement terms coming from the
second derivative of the pairing functional
$\mathcal{E}_{\mathrm{pair}}$ with respect to the normal density $\varrho$.

On the other hand, the residual interaction in the
particle-particle (p-p) channel is derived from the pairing
functional constructed with the density-dependent contact
interaction (\ref{pair_int}),
\begin{equation}
v_{pp}(\boldsymbol{r},\boldsymbol{r}^{\prime})=\dfrac{\delta^{2} \mathcal{E}_{\mathrm{pair}}}
{\delta \tilde{\varrho}(\boldsymbol{r}^{\prime}) \delta \tilde{\varrho}(\boldsymbol{r})}. \label{v_res_pp}
\end{equation}
This altogether coincides with Eq.~(\ref{pair_int}).

Because the full self-consistency between the static mean-field
calculation and the dynamical calculation is broken by the above
neglected terms, we renormalize the residual interaction in the
p-h channel by an overall factor $f_{ph}$ to get the spurious
$K^{\pi}=0^{-}$ and $1^{-}$  modes (representing the
center-of-mass motion), and $K^{\pi}=1^{+}$ mode (representing the
rotational motion in deformed nuclei) at zero energy ($v_{ph}
\rightarrow f_{ph}\cdot v_{ph}$). We cut the two-quasiparticle
(2qp) space at $E_{\alpha}+E_{\beta} \leq 60$ MeV due to the
excessively demanding computer memory size and computing time 
for the model space consistent with that adopted in the HFB
calculation. 
Accordingly, we need another factor $f_{pp}$ for the p-p channel. We determine
this factor such that the spurious $K^{\pi}=0^{+}$ mode associated
with the particle number fluctuation (representing the pairing
rotational mode) appears at zero energy ($v_{pp} \rightarrow
f_{pp}\cdot v_{pp}$).

\section{\label{results}Results and Discussion}
\subsection{$^{20}$O}
The application of our new calculation scheme is presented at
first for a spherical system.
In Ref.~\cite{miz07}, detailed properties of $^{20}$O were investigated using the
continuum QRPA based on the Skyrme density functional.
In the present subsection, we show our results for the isovector dipole
and isoscalar quadrupole modes in $^{20}$O following the
discussions in Ref.~\cite{miz07}.
In the HFB calculations, the pairing strengths $t_{0}^{\prime}=-280$ MeV$\cdot$fm$^{3}$
and $t_{3}^{\prime}=-18.75t_{0}^{\prime}$ with $\gamma=1$
are employed as in Ref.~\cite{miz07}.
With the choice $\gamma=1$ in the pairing interaction,
the pairing rearrangement terms vanish in the residual
interaction.

Because we use a larger mesh size and a smaller box,
the obtained solution is not exactly the same as in Ref.~\cite{miz07};
the calculated total binding energy is 157.7 MeV and
the average neutron pairing gap is
$\langle \Delta \rangle_{\nu}=1.92$ MeV.
There are also differences in the QRPA calculations between the two calculations: the
boundary conditions, the 2qp cutoff energy and the treatment of
the spin-dependent interactions ($\sigma \cdot \sigma^{\prime}$
terms in Eq.~(\ref{v_res_ph})). In the present calculation, the
transition spin density is treated exactly.

\begin{figure}[t]
\begin{center}
\includegraphics[scale=0.7]{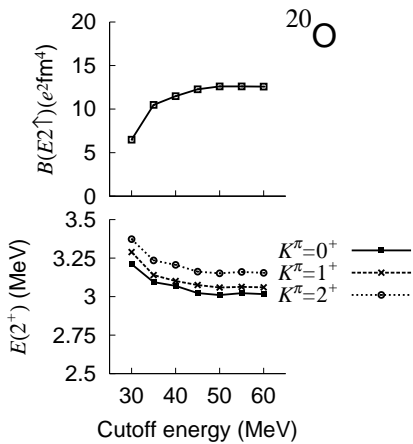}
\caption{The 2qp cutoff-energy dependence of the $B(E2\uparrow)$
values (upper panel) and excitation energies $E(2^{+})$
(lower panel) for the first excited $2^{+}$ state in $^{20}$O.}
\label{20O_dep}
\end{center}
\end{figure}

In Fig.~\ref{20O_dep} we show the 2qp cutoff energy dependence of
the excitation energies and electric quadrupole transition
strengths of the first excited state. In this figure, we show the
excitation energies of the $K^{\pi}=0^{+},1^{+}$ and $2^{+}$
states. If the spherical symmetry is preserved perfectly, these
energies should be degenerate. In the actual calculation, however,
the spherical symmetry is broken due to the finite mesh size.
Therefore, we can consider this difference ($\sim 150$ keV)
as the numerical error caused by the discretization of the
coordinates. The transition strength $B(E2\uparrow)$ is a sum of
the intrinsic transition strengths to the $K^{\pi}=0^{+}, \pm
1^{+}$ and $\pm 2^{+}$ states. Around the cutoff energy at 50 MeV,
one can see that both $E(2^{+})$ and $B(E2\uparrow)$ converge enough.

\begin{figure}[t]
\begin{center}
\includegraphics[scale=0.9]{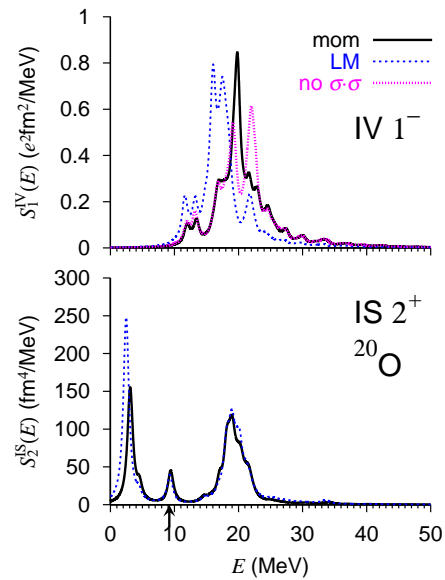}
\caption{Response functions for the isovector dipole (upper panel)
and the isoscalar quadrupole (lower panel) operators in $^{20}$O.
The dotted and solid lines correspond to the Landau-Migdal approximation (LM)
and the full calculation taken into account the momentum
dependence explicitly (mom). The dashed line denotes the
calculation without the $\sigma \cdot \sigma^\prime$
interaction while keeping the momentum dependence explicitly.
The transition strengths are smeared by a Lorentzian
function with a width of $\Gamma=1$ MeV. The arrow indicates the
neutron emission threshold $E_{\mathrm{th}}=9.19$ MeV.}
\label{20O_response}
\end{center}
\end{figure}

We show in Fig.~\ref{20O_response} the response functions for the isovector (IV) dipole
and the isoscalar (IS) quadrupole operators
\begin{subequations}
\begin{align}
\hat{F}_{1K}^{\mathrm{IV}}&=e\dfrac{N}{A}\sum_{i}^{Z}r_{i}Y_{1K}(\hat{r}_{i})-
e\dfrac{Z}{A}\sum_{i}^{N}r_{i}Y_{1K}(\hat{r}_{i}), \\
\hat{F}_{2K}^{\mathrm{IS}}&=\sum_{i}^{A}r^{2}_{i}Y_{2K}(\hat{r}_{i}),
\end{align}
\end{subequations}
and the corresponding response functions defined as
\begin{equation}
S_{\lambda}^{\tau}(E)
=\sum_{i}\sum_{K} \dfrac{\Gamma/2}{\pi}\dfrac{|\langle i|\hat{F}_{\lambda K}^{\tau}|0\rangle|^{2}}
{(E-\hbar \omega_{i})^{2}+\Gamma^{2}/4}.
\end{equation}
In this figure, we also show the response functions obtained by
using the LM approximation. This approximation treats only
approximately the momentum dependence of the Skyrme p-h residual
interaction. The resulting contact force is defined by the
density-dependent Landau parameters $F_{0},
F_{0}^{\prime}, G_{0}$ and $G_{0}^{\prime}$,
which are determined by the parameters of the Skyrme
effective interaction~\cite{gia81}.
The renormalization factors for the full QRPA calculation are
$f_{ph}=1.095$ and $f_{pp}=1.180$,
while $f_{ph}=0.815$ in the LM approximation.

For the IV dipole mode, the location of the giant resonance is
quite different between the calculation in the LM approximation
and that taking fully into account the momentum dependence.
The peak position is shifted up in the latter case.
In Ref.~\cite{miz07}, the same tendency was obtained.
However, the shape of the giant resonance is different between the two calculations.
This difference comes from the $\sigma \cdot \sigma^{\prime}$ terms
of the p-h interaction which were omitted in Ref.~\cite{miz07}.
Indeed, if we drop these terms in our calculation we obtain a two-peak structure
(see Fig.~\ref{20O_response}) which is consistent with the result of Ref.~\cite{miz07}.

For the IS quadrupole mode, we can see a prominent peak at $2-3$ MeV, as well as
the giant resonance at around 20 MeV. The low-lying state is sensitive to the
momentum-dependent components of the force while the
position of the giant resonance remains the same.
In both calculations we obtain a small peak at 9 MeV just above the threshold,
but the transition strength and the excitation energy are not affected.
This result is consistent with Ref.~\cite{miz07}.

\begin{figure}[t]
\begin{center}
\includegraphics[scale=0.9]{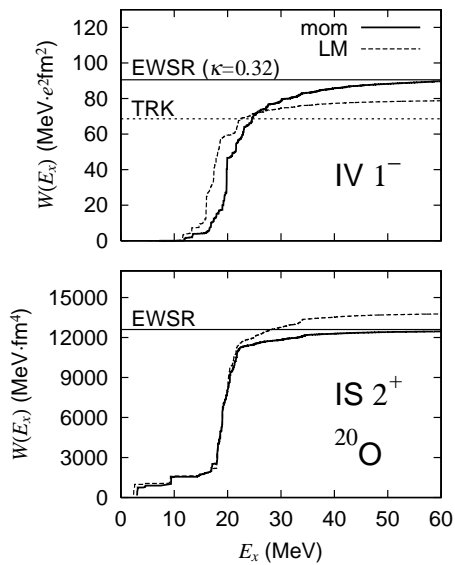}
\caption{Energy weighted sum of the IV dipole and IS quadrupole
strength functions. The solid and dotted lines represent the
calculation taking into account the momentum dependence explicitly
(mom) and in the Landau-Migdal approximation (LM), respectively.
The horizontal lines show the classical Thomas-Reiche-Kuhn (TRK), and
the RPA sum rule values for the isovector case including the enhancement factor,
$m_{1}=m_{1}^{\mathrm{cl}}(1+\kappa)$ ($\kappa=0.32$ in $^{20}$O
with the SkM* interaction ).
For the isoscalar case, the EWSR corresponds to the classical sum-rule value. }
\label{20O_EWSR}
\end{center}
\end{figure}

Figure~\ref{20O_EWSR} shows the partial sum of the energy weighted
strength defined as
\begin{equation}
W(E_{x})=\sum_{\hbar \omega_{i}<E_{x}}  \hbar \omega_{i}
|\langle i|\hat{F}_{\lambda}^{\tau}|0 \rangle|^{2}.
\end{equation}
For the IV dipole mode, the calculated sum up to 60 MeV
reaches 99.5\% of the EWSR value including the enhancement factor,
$m_{1}=m_{1}^{\mathrm{cl}}(1+\kappa)$~\cite{ter06} where
$\kappa=0.32$, whereas the calculation in the LM approximation
underestimates by 13\% the EWSR value.
For the IS quadrupole mode, the calculated sum satisfies 98.9\%
of the EWSR value,
which is comparable to the values obtained in Ref.~\cite{miz07}.
The calculation in the LM approximation overestimates
by 9.3\% the EWSR value.
This accuracy is as same as in Ref.~\cite{miz07}.

\begin{figure}[t]
\begin{center}
\includegraphics[scale=0.55]{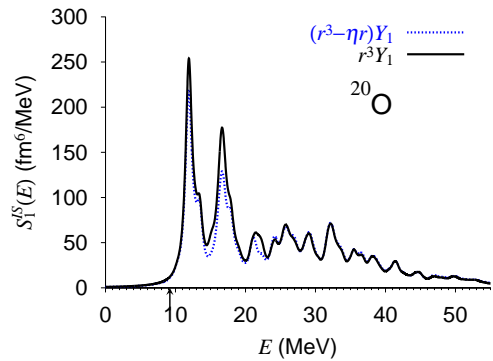}
\caption{Response functions for the IS dipole operators Eqs.~(\ref{IS_dip})
and (\ref{IS_dip_cor}) in $^{20}$O.
}
\label{20O_response_ISdip}
\end{center}
\end{figure}

Next, we discuss the decoupling of the spurious state from the physical states.
The IS dipole compression mode is
sensitive to the admixture of the center-of-mass
motion~\cite{ham02} because the response function to the IS dipole operator
\begin{equation}
\hat{F}_{1K}^{\mathrm{IS}}=\sum_{i}^{A}r^{3}_{i}Y_{1K}(\hat{r}_{i}) \label{IS_dip}
\end{equation}
contains the strengths of both the spurious mode and physical intrinsic excitations.
In order to see how much the spurious component mixes to the physical states,
we compare the response function to the operator Eq.~(\ref{IS_dip})
with that to the corrected operator~\cite{har81,gia81b}
\begin{equation}
\hat{F}_{1K}^{\mathrm{IS(cor)}}=\sum_{i}^{A}(r^{3}_{i}-\eta r_{i})Y_{1K}(\hat{r}_{i}),
\label{IS_dip_cor}
\end{equation}
where $\eta=\frac{3}{5}\langle r^{2} \rangle$.
If the spurious component is completely removed from the intrinsic
excitations, the calculated response functions to the operators
Eqs.~(\ref{IS_dip}) and (\ref{IS_dip_cor}) should coincide
with each other.

Figure~\ref{20O_response_ISdip} shows the response functions to the
IS dipole operators Eqs.~(\ref{IS_dip}) and (\ref{IS_dip_cor}).
At around 17 MeV, we can see a slight difference between the two responses.
In Ref.~\cite{ter05}, the admixture of the
spurious component was investigated in detail.
Accurate results could be obtained by using a very large cutoff energy
of 140 MeV in the 1qp spectrum included in the fully self-consistent QRPA calculations.
In the present work, the 2qp space is much smaller than in Ref.~\cite{ter05} and
some of the residual interactions are not included.
Though there is some room for further improvements,
the results obtained here are rather satisfactory.

\subsection{$^{26}$Ne}
In Ref.~\cite{yos08}, we have studied the properties of the
low-lying isovector resonance in deformed $^{26}$Ne using
the LM approximation. Although the strength function
observed at RIKEN~\cite{gib07} could be well reproduced,
the EWSR was not satisfied very accurately.
In the present subsection, we present the QRPA results
where the momentum dependence of $v_{ph}$ is fully included
and we show how the EWSR is fulfilled with the new method.
The parameters in the HFB calculation are the same as in Ref.~\cite{yos08},
the difference being only in the treatment of the momentum dependent
terms of interaction in the QRPA calculation.

\begin{figure}[t]
\begin{center}
\includegraphics[scale=0.6]{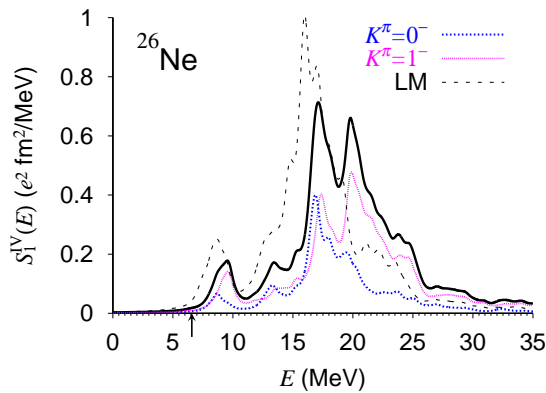}
\caption{Response functions for the isovector dipole operator in $^{26}$Ne.
The dotted, dashed and solid lines correspond to the $K^{\pi}=0^{-}$, $K^{\pi}=1^{-}$ and
total responses, respectively. The thin dotted line represents the response in the LM approximation. 
For the $K^{\pi}=1^{-}$ response, the transition strengths for the $K^{\pi}=\pm1^{-}$ states
are summed up.
The arrow indicates the neutron emission threshold $E_{\mathrm{th}}=6.58$ MeV.
}
\label{Ne_response}
\end{center}
\end{figure}

\begin{figure}[t]
\begin{center}
\includegraphics[scale=0.63]{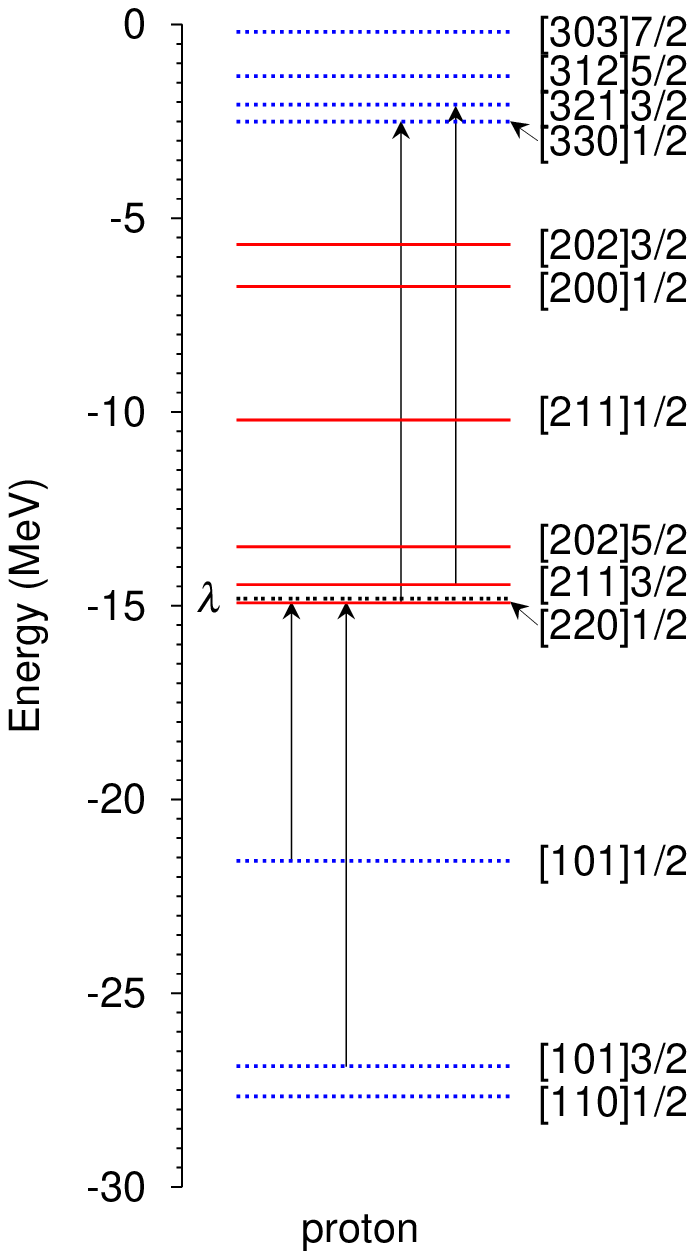}
\includegraphics[scale=0.63]{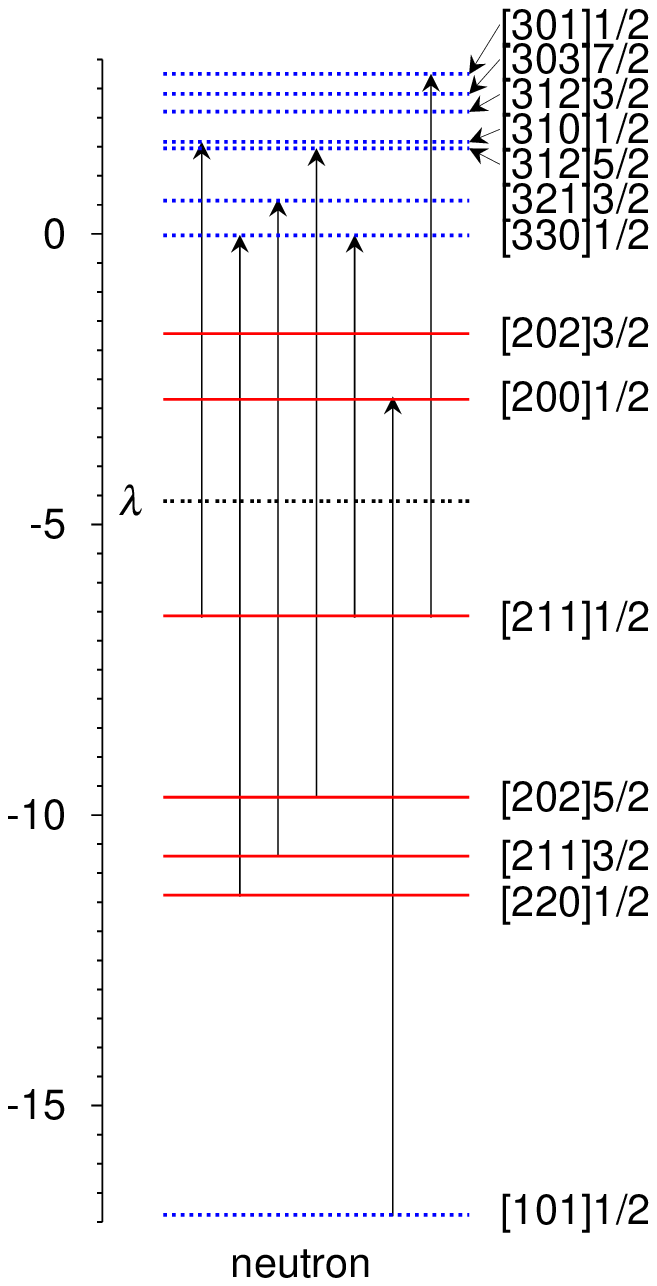}
\caption{ Two-quasiparticle (particle-hole)
excitations generating the $K^{\pi}=0^{-}$ state at 8.68 MeV in
$^{26}$Ne. The single-particle levels are labeled with the
asymptotic quantum numbers $[Nn_{3}\Lambda]\Omega$. 
The solid and dotted lines stand for the positive and negative
parities, respectively. The chemical potential $\lambda$ is
indicated by the two-dotted line.  }
\label{Ne_structure}
\end{center}
\end{figure}

\begin{table}[t]
\caption{QRPA amplitudes for the $K^{\pi}=0^{-}$ state at 8.68 MeV in $^{26}$Ne.
This mode has the isovector strength $B(Q^{\mathrm{IV}}1)=8.72 \times10^{-2} e^{2}$fm$^{2}$,
and the sum of backward-going amplitudes $\sum|Y_{\alpha\beta}|^{2}=5.24\times 10^{-3}$.
Only components with $X_{\alpha\beta}^{2}-Y_{\alpha\beta}^{2} > 0.001$ are listed.
In the rows (h) and (i), the label
$\nu 1/2^{-}$ denotes a non-resonant discretized continuum state of neutron
$\Omega^{\pi}=1/2^{-}$ level.
}
\label{26Ne_0-}
\begin{center}
\begin{tabular}{cccccc}
\hline \hline
 &  &  & $E_{\alpha}+E_{\beta}$ &  &
$Q_{10,\alpha\beta}$  \\
 & $\alpha$ & $\beta$ & (MeV) & $X_{\alpha \beta}^{2}-Y_{\alpha\beta}^{2}$ & ($e\cdot$ fm)
\\ \hline
(a) & $\nu[310]1/2$ & $\nu[211]1/2$ & 8.15 & 0.747 & $-0.309$ \\
(b) & $\nu[330]1/2$ & $\nu[220]1/2$ & 11.4 & 0.034 & $-0.397$ \\
(c) & $\nu[321]3/2$ & $\nu[211]3/2$ & 11.3 & 0.023 & 0.338 \\
(d) & $\nu[312]5/2$ & $\nu[202]5/2$ & 11.2 & 0.011 & $-0.239$ \\
(e) & $\nu[330]1/2$ & $\nu[211]1/2$ & 6.54 & 0.015 & $-0.118$ \\
(f) & $\nu[200]1/2$ & $\nu[101]1/2$ & 14.0 & 0.004 & $-0.241$ \\
(g) & $\nu[301]1/2$ & $\nu[211]1/2$ & 9.32 & 0.003 & $-0.117$ \\
(h) & $\nu 1/2^{-}$ & $\nu[211]1/2$ & 12.6 & 0.008 & $-0.068$ \\
(i) & $\nu 1/2^{-}$ & $\nu[211]1/2$ & 13.7 & 0.004 & $-0.077$ \\
\hline
(j) & $\pi[220]1/2$ & $\pi[101]1/2$ & 7.96 & 0.125 & 0.0085 \\
(k) & $\pi[220]1/2$ & $\pi[110]1/2$ & 14.1 & 0.014 & $-0.346$ \\
(l) & $\pi[330]1/2$ & $\pi[220]1/2$ & 13.4 & 0.013 & $-0.329$ \\
(m) & $\pi[321]3/2$ & $\pi[211]3/2$ & 14.0 & 0.004 & $-0.372$ \\
\hline \hline
\end{tabular}
\end{center}
\end{table}

\begin{figure}[t]
\begin{center}
\includegraphics[scale=0.57]{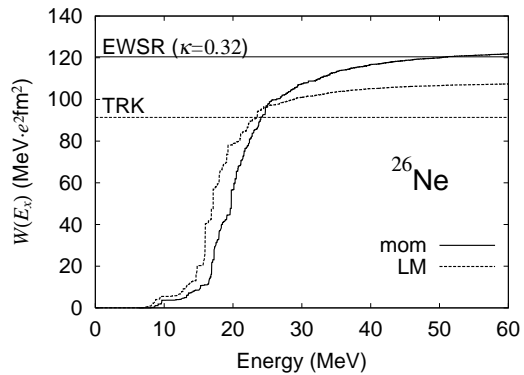}
\caption{Same as Fig.~\ref{20O_EWSR} but for $^{26}$Ne.
}
\label{26Ne_EWSR}
\end{center}
\end{figure}

We show in Fig.~\ref{Ne_response} the response functions for the
isovector dipole mode.
The renormalization factors for the QRPA calculation
are $f_{ph}=1.093$ and $f_{pp}=1.225$, while $f_{ph}=0.919$ in the LM approximation.
Compared to the response functions obtained
by using the LM approximation, the excitation energies of both the low-lying
and giant resonances are slightly shifted up
while the overall structure remains similar.
We can clearly see the resonance structure below 10 MeV as
experimentally observed~\cite{gib07}.
The resonance is governed by the $K^{\pi}=0^{-}$ state at 8.7 MeV and the $K^{\pi}=1^{-}$
states at 9.1 MeV and 9.6 MeV. The microscopic structure of the
$K^{\pi}=0^{-}$ state is given in
Fig.~\ref{Ne_structure} and Table~\ref{26Ne_0-}.
Here, the single-particle states are obtained by
rediagonalizing the self-consistent single-particle
Hamiltonian $h$ of Eq.~(\ref{HFB_equation}).
As in the case of the LM approximation, this $K^{\pi}=0^{-}$ state
contains mainly the neutron 1p-1h configuration $(2s_{1/2})^{-1}(2p_{3/2})$,
whose squared amplitude is 0.75.
The $K^{\pi}=1^{-}$ state at 9.1 MeV has also the same
main component, with a weight of 0.83.
The state at 9.6 MeV is mainly generated by the $(2s_{1/2})^{-1}(2p_{1/2})$ excitation
with a weight of 0.90.
A difference between the results of the LM approximation
and the present results is that
the transition strength to the $K^{\pi}=1^{-}$ state
at 9.6 MeV (0.08$e^{2}$fm$^{2}$) is larger than that
to the state at 9.1 MeV (0.04$e^{2}$fm$^{2}$).

Figure~\ref{26Ne_EWSR} shows the energy weighted sum of the
isovector dipole strength function together with the sum rule
values represented by the horizontal lines.
The energy-weighted sum up to 60 MeV overestimates by only 1.6\% the EWSR value
including the enhancement factor $\kappa=0.32$.
In the calculation with the LM approximation,
the energy-weighted sum is underestimated by about 10\%~\cite{yos08}.
This suggests that treating the momentum dependence of the Skyrme force explicitly
in the QRPA calculation is crucial for satisfying the EWSR in the
deformed systems as in the spherical systems~\cite{miz07}.
Because of the severe cutoff in the 2qp excitation energy, we cannot
describe properly the energy region higher than the giant resonance.
It would be interesting to see if calculations in a larger 2qp space including
the residual spin-orbit and the Coulomb interactions can improve the overshooting of the EWSR value.

\begin{figure}[t]
\begin{center}
\includegraphics[scale=0.6]{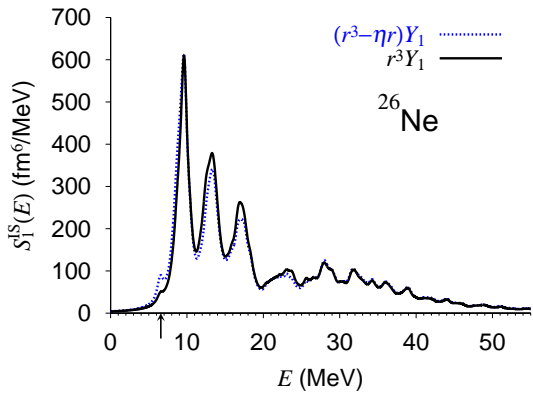}
\caption{Same as Fig.~\ref{20O_response_ISdip} but for $^{26}$Ne.
}
\label{26Ne_response_ISdip}
\end{center}
\end{figure}

Next, we discuss the IS dipole response. The corrected IS dipole
operator Eq.~(\ref{IS_dip_cor})
with $\eta=\frac{3}{5}\langle r^{2} \rangle$ is valid only for the
spherical systems. We can extend it for the deformed systems by
following the discussion in Appendix of Ref.~\cite{gia81b}.
One obtains the correction factor $\eta$ for the deformed systems
\begin{equation}
\eta=
\begin{cases}
3\langle z^{2} \rangle +\langle \rho^{2}\rangle & (K=0) \\
\langle z^{2}\rangle +2\langle \rho^{2}\rangle & (K=\pm 1).
\end{cases}
\end{equation}
These correction factors coincide with $\eta=\frac{3}{5}\langle r^{2} \rangle$ in the spherical limit.

In Fig.~\ref{26Ne_response_ISdip}, we show the response functions
to the IS dipole operators with/without the correction. For the
lowest state at 6.5 MeV, we can see a difference between the two
calculations. However, the overall structures are not very
different. We can consider that the spurious component is well
removed from the pygmy resonance and the giant resonance.

\subsection{$^{24}$Mg}
\begin{figure}[b]
\begin{center}
\includegraphics[scale=0.6]{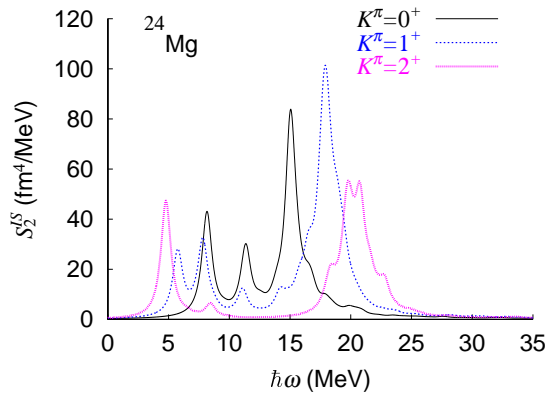}
\caption{Response functions for the IS quadrupole transition in $^{24}$Mg for
the $K^{\pi}=0^{+}, 1^{+}$ and $2^{+}$ excitations represented by the solid, dotted and dashed lines,
respectively.
}
\label{24Mg_response}
\end{center}
\end{figure}

We show in Fig.~\ref{24Mg_response}
the response functions for the IS quadrupole transition in $^{24}$Mg.
We employ the same effective interactions for the HFB+QRPA calculation as in the $^{20}$O nucleus.
The renormalization factor is $f_{ph}=1.164$.
The giant resonance appears at around 15 -- 25 MeV.
Since the ground state is prolately deformed in our calculation ($\beta_{2}=0.4$),
we can see a clear $K$-splitting.
Below 5 MeV, we can see a prominent peak for the $K^{\pi}=2^{+}$ excitation.
These are consistent with the fully self-consistent
deformed QRPA calculation using the Gogny force~\cite{per08}. 

\begin{figure}[t]
\begin{center}
\includegraphics[scale=0.55]{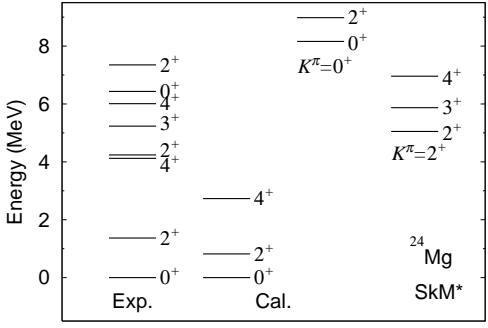}
\caption{Excitation energy spectrum obtained by the QRPA calculation
and the available experimental data~\cite{nds24}.
}
\label{24Mg_level}
\end{center}
\end{figure}

In Fig.~\ref{24Mg_level}, we show the low-lying excitation spectrum.
Here excitation energies are evaluated by~\cite{EG}
\begin{equation}
E(I,K)=\hbar \omega_{\mathrm{RPA}}+\frac{\hbar^{2}}{2\mathcal{J}_{\mathrm{TV}}}(I(I+1)-K^{2}),
\end{equation}
in terms of the vibrational frequencies, $\omega_{\mathrm{RPA}}$,
and the Thouless-Valatin moment of inertia~\cite{tho62},
$\mathcal{J}_{\mathrm{TV}}$, calculated microscopically by the
QRPA. The calculated moment of inertia is large compared to the
experimental rotational band. This is because the present
pairing interaction leads to vanishing neutron and proton pairing gaps. 
Furthermore, due to the absence of the coupling mechanism
between the $\beta$ vibration and the pairing vibration, the
excited $K^{\pi}=0^{+}$ mode cannot acquire a substantial 
collectivity and the excitation energy remains large. 
The $\gamma$ vibrational mode reasonably reproduces the experiment. 
This $K^{\pi}=2^{+}$ state is mainly generated 
by the neutron and proton p-h excitations $[211]3/2 \to [211]1/2$. 
Their contributions are 42\% and 56\%, respectively and the total
transition strength is exhausted about 55\% by the two p-h
excitations. The rest of the transition strength comes from the
coupling to the giant resonance. 

\subsection{$^{34}$Mg and $^{36}$Mg}
\begin{table}[b]
\caption{Ground state properties of $^{34}$Mg and $^{36}$Mg obtained by the
deformed HFB calculation with the SkM* interaction and the mixed-type pairing interaction.
Chemical potentials, deformations, average pairing gaps,
root-mean-square radii for neutrons and protons, and the total binding energies are listed.
}
\label{Mg_GS}
\begin{center}
\begin{tabular}{ccc} \hline \hline
 & $^{34}$Mg & $^{36}$Mg  \\ \hline
$\lambda_{\nu}$ (MeV) & $-4.16$ & $-3.24$  \\
$\lambda_{\pi}$ (MeV) & $-19.8$ & $-20.1$  \\
$\beta_{2}^{\nu}$ & 0.35 & 0.31  \\
$\beta_{2}^{\pi}$ & 0.41 & 0.39  \\
$\langle \Delta \rangle_{\nu}$ (MeV) & 1.71  & 1.71 \\
$\langle \Delta \rangle_{\pi}$ (MeV) & 0.0 & 0.0  \\
$\sqrt{\langle r^{2} \rangle_{\nu}}$ (fm) & 3.51 & 3.59  \\
$\sqrt{\langle r^{2} \rangle_{\pi}}$ (fm) & 3.16 & 3.18  \\
$E_{\mathrm{total}}$ (MeV) & 263.3 & 269.9 \\
\hline \hline
\end{tabular}
\end{center}
\end{table}

In Ref.~\cite{yos08b},
the properties of the low-lying
$K^{\pi}=0^{+}$ mode in $^{34}$Mg have been studied and the
moments of inertia were calculated using the Thouless-Valatin method.
In the present subsection,
effects of the momentum dependent components of the Skyrme interaction on
the isoscalar quadrupole strengths and the moments of inertia are discussed.

In the HFB calculations, the pairing strength parameter is
determined so as to reproduce the experimental pairing gap of
$^{34}$Mg ($\Delta_{\mathrm{exp}}=1.7$ MeV) obtained by the
three-point formula~\cite{sat98}. 
In Table~\ref{Mg_GS}, the
ground state properties of $^{34,36}$Mg are summarized. 
The strength $t_{0}^{\prime}=-295$MeV$\cdot$fm$^{3}$ for the
mixed-type interaction with $\gamma=1$ leads to the pairing gaps
$\langle \Delta_{\nu}\rangle=1.71$ MeV in $^{34}$Mg and $^{36}$Mg.
We obtain for the proton intrinsic quadrupole moments $Q_{0}$ the
values 62.2$e\cdot$fm$^{2}$ and 60.1$e\cdot$fm$^{2}$ in $^{34}$Mg
and $^{36}$Mg, respectively. The reduced transition
probabilities~\cite{RS} are then $B(E2;0^{+}\to 2_{1}^{+})=5/16\pi
\cdot Q_{0}^{2} =385 e^{2}\cdot$fm$^{4}$ and 359
$e^{2}\cdot$fm$^{4}$ in $^{34}$Mg and $^{36}$Mg. 
In $^{34}$Mg, the neutron occupation probability of the [202]3/2
level coming up from the $1d_{3/2}$ orbit is 0.28, while that of
the [321]3/2 level coming down from the $1f_{7/2}$ orbit is 0.67.
This approximately corresponds to the
$(1f_{7/2})^{4}(1d_{5/2})^{-2}$ configuration in the spherical
shell model language. In $^{36}$Mg, the occupation probability of
the $\nu[202]3/2$ level becomes 0.64. 
These probabilities are consistent with the shell model results~\cite{uts99}. 

In the QRPA calculations of $^{34}$Mg and $^{36}$Mg,
we cut the 2qp space at 60 MeV as in the previous subsections.
We have checked that the transition strength to the excited
$0^{+}$ state and its excitation energy converge at 50 MeV cutoff.
In the present case, the dimension of the QRPA
matrix~(\ref{eq:AB1}) for the $K^{\pi}=0^{+}$ channel in $^{36}$Mg
is about 15,000, the memory size is 20.8 GB,
and the CPU time is about 154,000s  per iteration for determining the
renormalization factor $f_{pp}$ using the SX-8 supercomputer at the RCNP.

\begin{figure}[t]
\begin{center}
\includegraphics[scale=0.85]{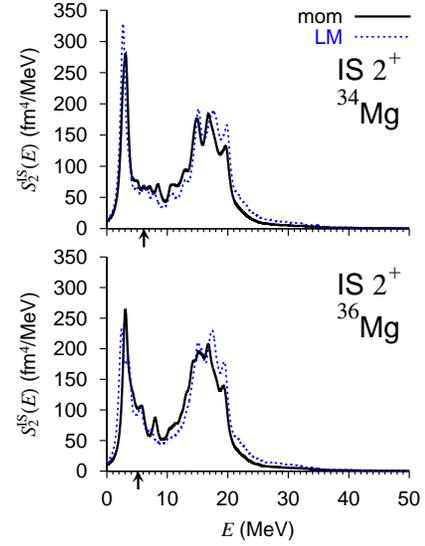}
\caption{Response functions for the IS quadrupole operator in $^{34}$Mg (upper panel)
and $^{36}$Mg (lower panel).
The arrows indicate the neutron emission thresholds $E_{\mathrm{th}}=6.13$ MeV and 5.18 MeV
for $^{34}$Mg and $^{36}$Mg, respectively.}
\label{Mg_response}
\end{center}
\end{figure}

Figure~\ref{Mg_response} shows the response functions for the IS quadrupole operator
in $^{34}$Mg and $^{36}$Mg.
The renormalization factors for $^{34}$Mg and $^{36}$Mg are
$f_{ph}=1.146$ and 1.139, respectively.
In the pp channel, the renormalization factors are
$f_{pp}=1.197$ and 1.215 for $^{34}$Mg and $^{36}$Mg.

In both nuclei, we can see a low-lying peak at 2 MeV and a three-peak giant resonance
at $15-20$ MeV. This three-peak structure corresponds to the giant resonance
for the $K^{\pi}=0^{+}, 1^{+}$ and $2^{+}$ excitations.
Because the deformation of $^{34}$Mg is larger than that of $^{36}$Mg,
the $K$ splitting is larger in $^{34}$Mg.
On the same figure are shown for comparison the results of the LM approximation.
As in $^{20}$O, the low-lying state is
sensitive to the treatment of the momentum-dependent interactions
while the position of the giant resonance is not much affected.

The calculated energy-weighted sums up to 60 MeV
for the $K^{\pi}=0^{+}$ excitation in $^{34}$Mg and $^{36}$Mg amount to 99.0\%,
whereas in the LM approximation, they are overestimated by
11.5\% and 11.0\% the EWSR in $^{34}$Mg and $^{36}$Mg.
This confirms that the EWSR is well satisfied in the present calculation.

\begin{figure}[t]
  \begin{center}
    \begin{tabular}{ccc}
    \includegraphics[height=4.7cm]{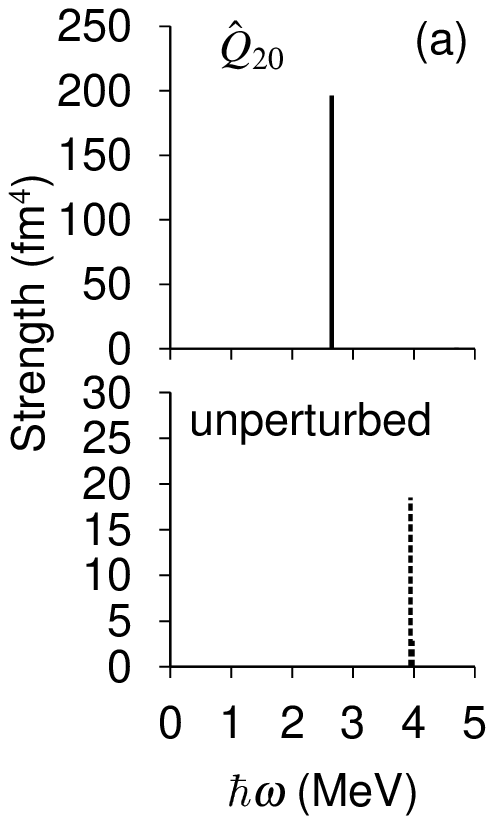}
    \includegraphics[height=4.7cm]{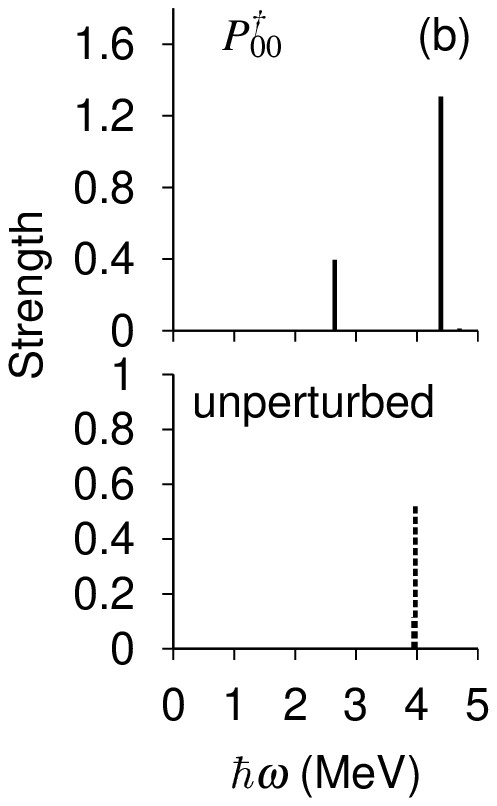}
    \includegraphics[height=4.7cm]{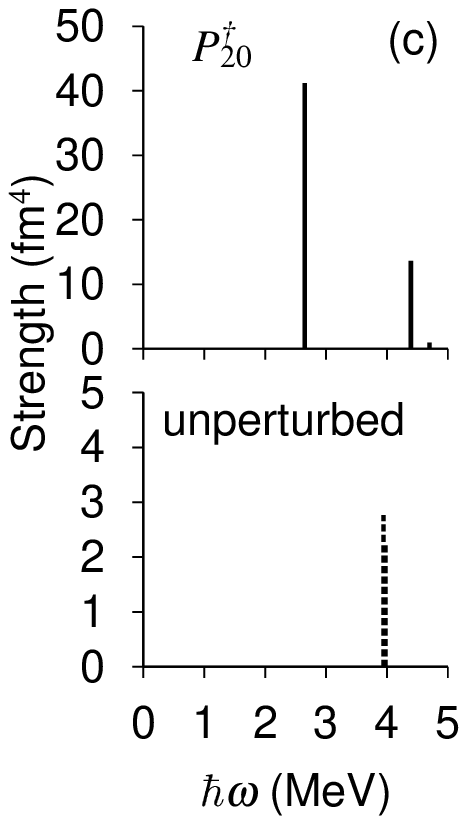}
    \end{tabular}
\caption{QRPA strength distributions for
(a) the isoscalar $K^{\pi}=0^{+}$ quadrupole p-h excitations
and (b) the monopole- and
(c) the quadrupole-pair excitations in $^{34}$Mg.
For comparison, unperturbed 2qp transition strengths are shown
in the lower panels.}
\label{34Mg_strength}
\end{center}
\end{figure}

In Ref.~\cite{yos08b}, we have also discussed
the generic feature of the low-lying $K^{\pi}=0^{+}$ modes in deformed
neutron-rich nuclei:
In a deformed system where the up-sloping
oblate-type and the down-sloping prolate-type orbitals exist near
the Fermi level, one obtains a low-lying mode possessing enhanced
strengths both for the quadrupole p-h transition and for the
quadrupole p-p (pair) transition induced by the pairing
fluctuations. In Fig.~\ref{34Mg_strength}, we show the strength
distributions in $^{34}$Mg of the quadrupole p-h, the
monopole p-p and the quadrupole p-p transitions defined by the operators
\begin{subequations}
\begin{align}
\hat{Q}_{20}&=\sum_{q,\sigma}\int d\boldsymbol{r} r^{2}Y_{20}(\hat{r})
\hat{\psi}^{\dagger}_{q}(\boldsymbol{r}\sigma)\hat{\psi}_{q}(\boldsymbol{r}\sigma), \\
\hat{P}^{\dagger}_{00}&=\int d\boldsymbol{r}
\hat{\psi}^{\dagger}_{\nu}(\boldsymbol{r}\uparrow)\hat{\psi}^{\dagger}_{\nu}(\boldsymbol{r}\downarrow), \\
\hat{P}^{\dagger}_{20}&=\int d\boldsymbol{r} r^{2}Y_{20}(\hat{r})
\hat{\psi}^{\dagger}_{\nu}(\boldsymbol{r}\uparrow)\hat{\psi}^{\dagger}_{\nu}(\boldsymbol{r}\downarrow).
\end{align} \label{eq:pair_transition}
\end{subequations}

At 2.65 MeV, we obtain the collective $K^{\pi}=0^{+}$ mode
possessing about 30 Weisskopf units for the intrinsic isoscalar
quadrupole transition strength. (The electric transition strength
is $B(E2;0_{2}^{+} \to 2_{1}^{+})=13.0 e^{2}$fm$^{4}$.)
The transition strength is enhanced
by 10.6 times as compared to the unperturbed transition strength.
For the quadrupole pair transition, the strength to this
collective state is enhanced by 14.9 times with respect
to the unperturbed one, while the strength is not changed for the
monopole pair transition.
We have checked that the low-lying
$K^{\pi}=0^{+}$ mode at 2.65 MeV is well decoupled from the pairing rotation.
The transition strength for the number operator
to this state is $|\langle \lambda |\hat{N}|0\rangle|^{2}=2.87 \times 10^{-5}$.

\begin{figure}[t]
\begin{center}
\includegraphics[height=4.0cm]{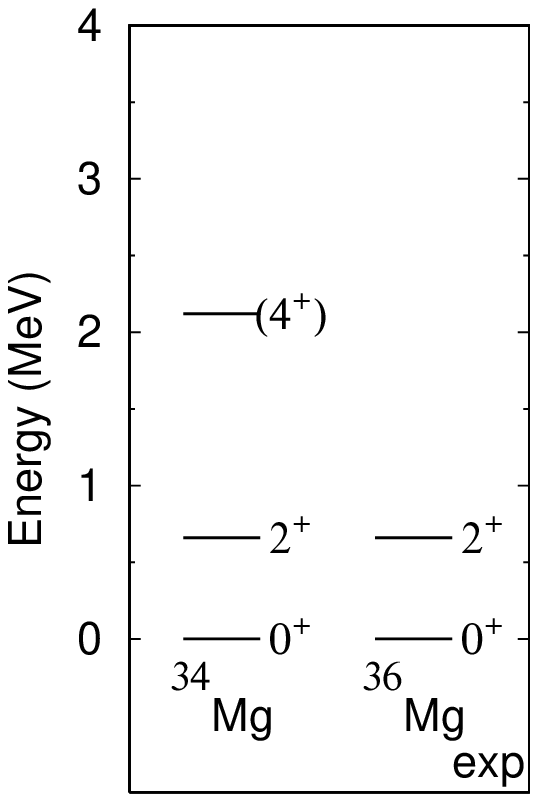}
\includegraphics[height=4.0cm]{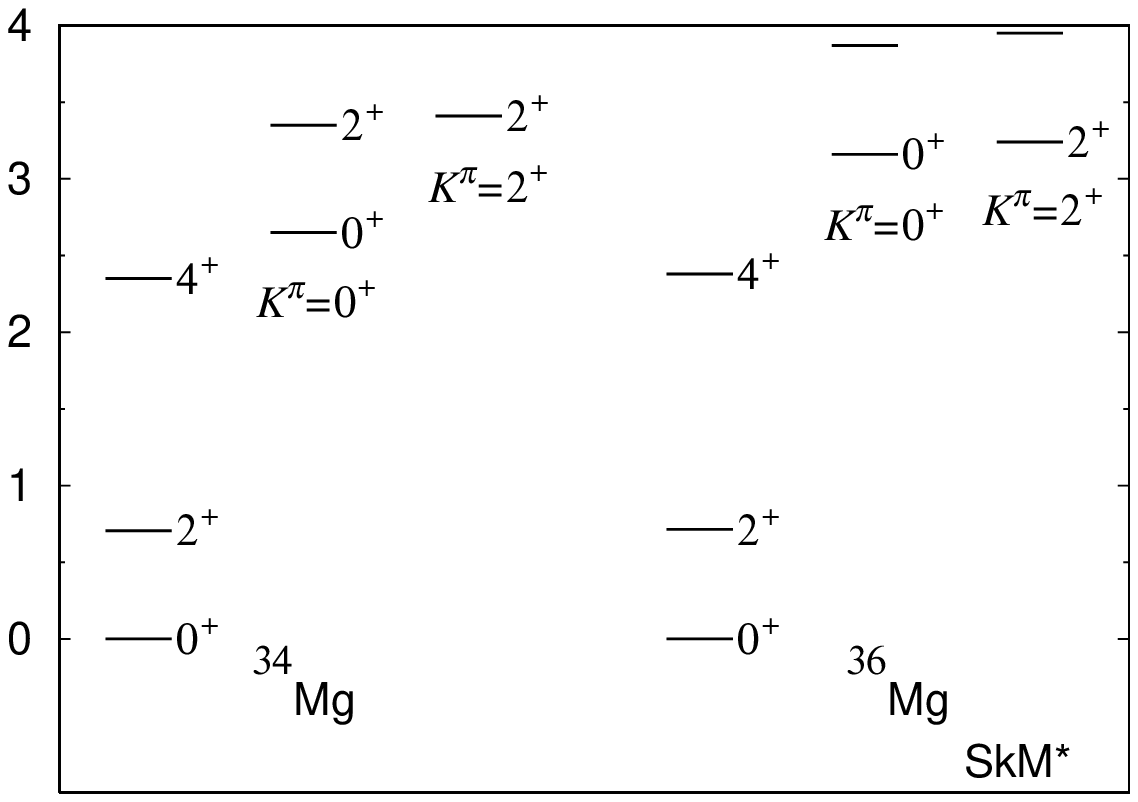}
\caption{Excitation energy spectrum for the positive parity states in $^{34}$Mg and $^{36}$Mg,
and the available experimental data~\cite{yon99,gad07}. }
\label{Mg_level}
\end{center}
\end{figure}

In Fig.~\ref{Mg_level}, we show the low-lying excitation spectrum
below 4 MeV for the positive parity states in $^{34}$Mg and $^{36}$Mg.
The Thouless-Valatin moment of inertia of $^{34}$Mg is
$\mathcal{J}_{\mathrm{TV}}/\hbar^{2}=4.23$ MeV$^{-1}$
when the LM approximation is used,
and 4.26 MeV$^{-1}$ when the full momentum-dependent interaction
is treated in the QRPA,
while the Inglis-Belyaev moment of inertia is
$\mathcal{J}_{\mathrm{Belyaev}}=3.89$ MeV$^{-1}$.
Due to the time odd components in the residual interactions
(\ref{v_res_ph}) and (\ref{v_res_pp}),
the moment of inertia $\mathcal{J}_{\mathrm{TV}}$ becomes about 10 \% larger than
$\mathcal{J}_{\mathrm{Belyaev}}$.
In $^{36}$Mg, we obtain
$\mathcal{J}_{\mathrm{TV}}/\hbar^{2}=4.20$ MeV$^{-1}$,
and 4.24 MeV$^{-1}$ in the LM approximation.
If we turn off the residual interactions we obtain 3.84 MeV$^{-1}$.
For both nuclei, the
moments of inertia calculated by using the LM approximation are
close to the results of the QRPA with the full velocity-dependent force.

\section{\label{summary}Conclusions}
We have developed a new calculation scheme of the deformed QRPA
using the Skyrme density functional.
This scheme allows one to include in the p-h residual
interaction the full velocity dependence of the Skyrme force.
The only components of $v_{ph}$ not treated here are the spin-orbit
and Coulomb two-body forces.
Numerical applications have been performed
for some spherical and deformed neutron-rich nuclei
employing the Skyrme SkM* density functional and
a local pairing functional for generating the HFB mean field, pairing field
and the residual interactions in the QRPA calculations.
In both the spherical and the deformed cases,
we have checked that the energy-weighted sum rules for
the isoscalar and the isovector operators are well satisfied.
There is a distinct improvement over the sum rules predicted by the LM
approximation, even in the case of isoscalar excitations, thus
indicating the importance of full self-consistency in HF(B)-(Q)RPA calculations.

Thus, this method enables one to describe multipole
strengths quantitatively in deformed nuclei located in a wide range of the nuclear
chart, even near drip lines.
It has been also shown that
one can apply it not only to the vibrational modes but also to rotational modes by
employing the Thouless-Valatin procedure.

Methods for solving the HFB-QRPA problem in deformed systems using
the Skyrme plus local pairing density functional in a fully
consistent way are still scarce. We have proposed here a new
method and we have demonstrated its feasibility on some examples.
This method has some advantages and drawbacks. One main advantage
is the choice of solving the deformed HFB problem on a grid in
coordinate space. This avoids expanding quasiparticle wave
functions on a harmonic oscillator basis and introducing
inaccuracies inherent to expansions of loosely bound, or unbound
wave functions on such basis. This may be of some importance when
studying near drip-line nuclei. A practical drawback is the
necessity of using a relatively large 2qp cutoff, and therefore
computing times and memory storage are high. Our numerical studies
show that a rather good accuracy is already reached if the 2qp
energy cutoff is set at 60 MeV. Doubtless that, in the future the
capacity of computing facilities will largely improve and the 2qp
space can be easily enlarged.

\begin{acknowledgments}
One of the authors (K.Y) is supported by the Special Postdoctoral Researcher Program of RIKEN.
This work was supported by the JSPS Core-to-Core Program
``International Research Network for Exotic Femto Systems".
The numerical calculations were performed on the NEC SX-8 supercomputers
at Yukawa Institute for Theoretical Physics, Kyoto University and
at Research Center for Nuclear Physics, Osaka University.
\end{acknowledgments}

\begin{appendix}
\section{The Skyrme density functional}
The total energy of the system consists of the kinetic energy
$\mathcal{E}_{\mathrm{kin}}$, the Skyrme interaction energy
$\mathcal{E}_{\mathrm{Sky}}$, the Coulomb energy
$\mathcal{E}_{\mathrm{Coul}}$, the pairing energy
$\mathcal{E}_{\mathrm{pair}}$ and the correction of center-of-mass
motion and rotational motion $\mathcal{E}_{\mathrm{corr}}$;
\begin{equation}
\mathcal{E}=\mathcal{E}_{\mathrm{kin}}+\mathcal{E}_{\mathrm{Sky}}
+\mathcal{E}_{\mathrm{Coul}}+\mathcal{E}_{\mathrm{pair}}+\mathcal{E}_{\mathrm{corr}},
\end{equation}
The kinetic energy is given by
\begin{equation}
\mathcal{E}_{\mathrm{kin}}=\int d\boldsymbol{r} \dfrac{\hbar^{2}}{2m}\tau(\boldsymbol{r}),
\end{equation}
where $\tau$ is the kinetic density.
In the present paper, we perform the numerical calculations using
the SkM* interaction~\cite{bar82},
so the center-of-mass correction is just to replace the nucleon mass $1/m \to 1/m \times (1-1/A)$.
The correction for the rotational motion is not taken into account.

The Skyrme interaction energy is given as~\cite{eng75,dob95}
\begin{equation}
\mathcal{E}_{\mathrm{Sky}}=\int d\boldsymbol{r} \mathcal{H}_{\mathrm{Sky}}(\boldsymbol{r}),
\end{equation}
\begin{align}
\mathcal{H}_{\mathrm{Sky}}(\boldsymbol{r})
=& \sum_{t=0,1}\Bigl\{
C_{t}^{\rho}[\varrho_{0}(\boldsymbol{r})]\varrho^{2}_{t}(\boldsymbol{r})
+ C_{t}^{\boldsymbol{s}}[\varrho_{0}(\boldsymbol{r})]\boldsymbol{s}^{2}_{t}(\boldsymbol{r}) \notag \\
&+ C_{t}^{\triangle \rho}\varrho_{t}(\boldsymbol{r})\triangle \varrho_{t}(\boldsymbol{r})
+ C_{t}^{\triangle \boldsymbol{s}}\boldsymbol{s}_{t}(\boldsymbol{r})
\cdot \triangle \boldsymbol{s}_{t}(\boldsymbol{r}) \notag \\
& + C_{t}^{\tau}(\varrho_{t}(\boldsymbol{r})\tau_{t}(\boldsymbol{r})-\boldsymbol{j}^{2}_{t}(\boldsymbol{r})) \notag \\
& + C_{t}^{T}(\boldsymbol{s}_{t}(\boldsymbol{r})\cdot \boldsymbol{T}_{t}(\boldsymbol{r})
-\overleftrightarrow{J}^{2}_{t}(\boldsymbol{r})) \notag \\
&+ C_{t}^{\nabla J}(\varrho_{t}(\boldsymbol{r})\nabla \cdot \boldsymbol{J}_{t}(\boldsymbol{r})
+\boldsymbol{s}_{t}(\boldsymbol{r})\cdot \nabla \times \boldsymbol{j}_{t}(\boldsymbol{r})) \Bigr\}, \label{Sky_energy}
\end{align}
where $\varrho$ denotes the nucleon density, $\boldsymbol{s}$ the
spin density, $\boldsymbol{T}$ the kinetic spin density,
$\boldsymbol{j}$ the current tensor, $\overleftrightarrow{J}$ the
spin-current tensor and $\boldsymbol{J}$ the spin-orbit current.
All densities are labelled by an isospin index $t$ where $t$ 
is 0 (isoscalar) or 1 (isovector) and we assume no isospin mixing.

The Coulomb energy is given as
\begin{align}
\mathcal{E}_{\mathrm{Coul}}&=\int d\boldsymbol{r} \mathcal{H}_{\mathrm{Coul}}(\boldsymbol{r}), \notag \\
\mathcal{H}_{\mathrm{Coul}}(\boldsymbol{r})&=\dfrac{e^{2}}{2}\int d\boldsymbol{r}^{\prime}
\varrho_{\pi}(\boldsymbol{r})\dfrac{\varrho_{\pi}(\boldsymbol{r}^{\prime})}
{|\boldsymbol{r}-\boldsymbol{r}^{\prime}|}
-\dfrac{3e^{2}}{4}\left(\dfrac{3}{\pi} \right)^{\frac{1}{3}} \varrho_{\pi}^{4/3}(\boldsymbol{r}),
\end{align}
where the exchange term in the Coulomb energy is treated in the
Slater approximation~\cite{sla51}, and the higher order correction
is found to be small~\cite{tit74}. We follow the procedure of
Ref.~\cite{vau73} for calculating the Coulomb potential.

When we use for the pairing interaction the form of Eq.~(\ref{pair_int}),
the pairing energy is given as
\begin{align}
\mathcal{E}_{\mathrm{pair}}&=\int d\boldsymbol{r} \mathcal{H}_{\mathrm{pair}}(\boldsymbol{r}), \notag \\
\mathcal{H}_{\mathrm{pair}}(\boldsymbol{r})&=\dfrac{1}{8}
\left[t_{0}^{\prime}+\dfrac{t_{3}^{\prime}}{6}\varrho_{0}^{\gamma}(\boldsymbol{r}) \right]
\sum_{t=0,1}(\tilde{\varrho}_{t}^{2}(\boldsymbol{r})-\tilde{\boldsymbol{s}}_{t}^{2}(\boldsymbol{r})),
\end{align}
where $\tilde{\varrho}$ denotes the abnormal (pairing) density and $\tilde{\boldsymbol{s}}$
the spin pairing density.

\end{appendix}

\end{document}